\documentclass[aoas,preprint]{imsart}

\usepackage[OT1]{fontenc}
\usepackage{amsmath,mathrsfs,amssymb,amsfonts,mathrsfs}
\usepackage{booktabs,threeparttable,multirow}
\usepackage{bm,supertabular}

\usepackage{amssymb}
\usepackage{amsmath}
\usepackage{enumerate}
\usepackage{booktabs}
\usepackage{multirow}
\usepackage{bm}
\usepackage{latexsym}
\usepackage{graphicx}
\usepackage{subfigure}
\usepackage{multirow,booktabs}
\usepackage{indentfirst}
\usepackage{color}
\usepackage{url}
\usepackage{caption}

\RequirePackage[OT1]{fontenc}
\RequirePackage{amsthm,amsmath,mathrsfs,amsfonts,amssymb}
\RequirePackage{booktabs,threeparttable,tabularx,multirow}
\RequirePackage{bm,supertabular}
\RequirePackage{caption,url}
\RequirePackage{natbib}
\RequirePackage[colorlinks,citecolor=blue,urlcolor=blue]{hyperref}

\arxiv{arXiv:0000.0000}

\startlocaldefs
\numberwithin{equation}{section}
\theoremstyle{plain}

\endlocaldefs

\begin{document}
 \renewcommand\arraystretch{1.5}

\begin{frontmatter}
\title{Main and Interaction Effects Selection for Quadratic Discriminant Analysis via Penalized Linear Regression\thanksref{T1}}
\runtitle{Main and Interaction Effects Selection for QDA}
\thankstext{T1}{This work is partially
supported by the National Science Foundation of China (11101005, 11571021).}

\begin{aug}
\author{\fnms{Deqiang} \snm{Zheng}\thanksref{m1,m2,m3}\ead[label=e1]{dqzheng@ccmu.edu.cn}},
\author{\fnms{Jinzhu} \snm{Jia}\thanksref{m2,t1}\ead[label=e2]{jzjia@math.pku.edu.cn}}
\and
\author{\fnms{Xiangzhong} \snm{Fang}\thanksref{m2}
\ead[label=e3]{xzfang@math.pku.edu.cn}}
\and
\author{\fnms{Xiuhua} \snm{Guo}\thanksref{m1,m3}
\ead[label=e4]{guoxiuh@ccmu.edu.cn}}

\thankstext{t1}{Correspondence to: jzjia@math.pku.edu.cn.}
\runauthor{D. Zheng et al.}

\affiliation{School of Public Health, Capital Medical University\thanksmark{m1}\\
School of Mathematical Sciences, Peking University\thanksmark{m2}\\
Beijing Municipal Key Laboratory of Clinical Epidemiology\thanksmark{m3}\\
}

\end{aug}

\begin{abstract}

Discriminant analysis is a useful classification
method. Variable selection for discriminant analysis
is becoming more and more important in a high-dimensional setting.
This paper is concerned with the binary-class problems of main and
interaction effects selection for the quadratic discriminant analysis.
We propose a new penalized quadratic discriminant analysis (QDA)
for variable selection in binary classification.
Under sparsity assumption on the relevant variables,
we conduct a penalized liner regression
to derive sparse QDA by plugging the main and interaction effects in the model. Then the QDA problem is converted to a
penalized sparse ordinary least squares optimization
by using the composite absolute penalties (CAP).
Coordinate descent algorithm is introduced to solve the
convex penalized least squares. The penalized linear regression
can simultaneously select the main and interaction effects, and also
conduct classification. Compared with the existing methods
of variable selection in QDA,
the extensive simulation studies and two real data analyses
demonstrate that our proposed method works well and
is robust in the performance of variable selection and classification.
\end{abstract}



\end{frontmatter}

\section{Introduction}
\label{s:intro}

Nowadays supervised classification has been an important problem in
various medical fields such as genomic, disease diagnosis and brain imaging.
Many classification methods have been developed,
including linear and quadratic discriminant analysis (LDA and QDA)
 \citep{Anderson1984},
 k-nearest-neighbors \citep{Fix1951}, logistic regression \citep{Cox1958}, classification tree \citep{Breiman1984} and SVM \citep{Boser1992}.
The referred methods above are introduced and summarized
in the book \citep{Hastie2009}.
 Among many classification
methods, discriminant analysis is widely used in many applications due to simplicity, interpretability, and effectiveness. In many cases, it is believed that
only a subset of the available variables (also be called features or predictors)
may be contained in the classification structure (or model).
When irrelevant predictors are added into the model,
they may bring in extra noise, and the classification
performance may be degraded due to the unstable and inaccurate
estimations of the parameters.
Therefore, conducting variable selection before fitting the
model is advisable.
Variable selection can identify fewer discriminative variables
and  provide a more accurate classification model to describe
the future data.
Model selection methods are usually used to carry out the variable selection
in a probabilistic framework.

A BIC-type criterion for variable selection on quadratic discriminant analysis has been
recently studied by \citet{Zhang2011} and \citet{Murphy2010}.
The BIC-type model assumes that the relevant variables and
irrelevant variables jointly
follow a multivariate normal distribution.
The relevant variables have different means or covariances in different classes,
and irrelevant variables are conditionally independent of the class label.
It means that the irrelevant variables can be completely modeled by
a multivariate normal distribution conditionally on the relevant variables.
The BIC criteria are based on the full
likelihood of mixtures of multivariate normal distributions.
\citet{Zhang2011} proposed a standard backward algorithm
to find the set of relevant variables, and \citet{Murphy2010} used a
forward-backward algorithm for the variable selection.
\citet{Zhang2011} also showed the BIC's selection
consistency under the normal assumption.
However, performance may be compromised when this normal assumption does not hold.
Moreover, LDA and QDA are inapplicable for
the high-dimensional cases when the model dimensionality $p$
exceeds the sample size $n$, since the sample covariance matrices are consequently singular.

Lasso-type regularization methods \citep{Tibshirani1996,Zhao2006}
are popular in the literature
for high-dimensional variable selection. The Lasso-type regularization procedures impose constraints represented
by a penalty function, among which
$L_1$-norm and $L_2$-norm penalties have been previously explored
for variable selection.
\citet{Fan2010} provided a good review on variable selection and penalty functions.
In the high-dimensional classification literature, the Lasso-type regularization methods
have been frequently used for variable selection.
Among them, \citet{Cai2011} proposed a direct approach to sparse LDA by estimating
the product of precision matrix and the mean vector of two classes,
and \citet{Mai2012} also introduced a direct approach to
transform the LDA problem to a penalized linear regression.
\citet{Fan2015} proposed a two-step procedure to sparse QDA (IIS-SQDA),
where an innovated interaction screening approach was
explored based on the innovated transform of the precision matrices
of two classes in the first step and
a sparse quadratic discriminant analysis
was presented for further selecting important interactions and main effects
and conducting classification simultaneously in the second step.
\citet{Fan2015} also proved the consistency
of the estimated coefficient vector of QDA, and further showed that the classification error of IIS-SQDA could be infinitely close to the oracle classification
error. However, IIS-SQDA is based on the assumption that
the variables follow a Gaussian mixture distribution with conditional independence.
If the relevant predictors do not follow the normal assumption,
many irrelevant predictors can be selected. Even if the relevant predictors and irrelevant predictors are discriminated correctly,
the performance of classification may be much compromised.

In this work we consider binary classification problem with possibly unequal
means or covariance matrices.
Under some sparsity assumption on the relevant variables, we suggest using the penalized liner regression to
derive sparse QDA by plugging the main and interaction effects in the model. Motivated by the sparse LDA approach
explored by the method of sparse LDA in \citet{Mai2012}, we transform the QDA problem to a
penalized sparse ordinary least squares optimization.
We intuitively suppose that an
interaction effect should be added to the regression model only after the corresponding main effects. Therefore, we propose using the composite absolute penalties (CAP)
which was introduced by \citet{Zhao2009}.
Coordinate descent algorithm is presented to solve the
convex penalized least squares. The penalized linear regression
can simultaneously select the main and interaction effects, and also
conduct classification. Extensive simulation studies and real data analysis
demonstrate that our proposed method works well and is
more robust than the existing methods in both
the performance of variable selection
and classification error.

The rest of the paper is organized as follows.
Section $2$ introduces the
discriminant analysis and existing variable selection methods.
Section $3$ proposes the penalized
linear regression of sparse quadratic discriminant analysis.
The penalized linear regression is established, where the composite absolute penalty
is used to carry out variable selection. The coordinate descent algorithm is
presented to solve the penalized least squares optimization.
Extensive  simulation studies and applications to two real data examples
are presented in Section $4$ and Section $5$, respectively. Section $6$
concludes with a discussion.

\section{Discriminant analysis and existing variable selection methods}
\label{s:model}
We consider a binary classification problem.
Let $X\in\mathbb{R}^p$
be a vector of $p$ continuous predictor variables
and $G\in\{1,2\}$ represents the class label.
The quadratic discriminate analysis assumes that $P(G=k)=\pi_k>0$
for $k=1,2$ and $X|G=k$ follows a multivariate normal distribution $N(\mu_k,\Sigma_k),k=1,2$.
Here $\mu_k=(\mu_{k1},\mu_{k2},\cdots,\mu_{kp})^T\in\mathbb{R}^p$
and $\Sigma_k\in\mathbb{R}^{p\times p}$ denote the mean vector and
covariance matrix for the predictors $X$ in the $k$-th class,respectively.
Then the quadratic discriminant function is
\begin{equation*}
    \delta_k(x|\pi_k,\mu_k,\Sigma_k)  =  -\frac{1}{2}\log \det(\Sigma_k)-\frac{1}{2}(x-\mu_k)^T\Sigma_k^{-1}(x-\mu_k)+\log\pi_k,k=1,2
\end{equation*}
where $x\in\mathbb{R}^{p}$ is the
column vector of the predictors for one observation.
Let $\hat\pi_k$, $\hat\mu_k$ and $\hat\Sigma_k$
be the estimates of $\pi_k$, $\mu_k$ and $\Sigma_k$.
Then the optimal Bayes rule minimizing is to
predict the new subject as the class with the maximal discriminant function value,
\begin{equation*}
    \hat G = \mathop{\arg\max}_{k}\delta_k(x|\hat\pi_k,\hat\mu_k,\hat\Sigma_k),k=1,2.
\end{equation*}

Recently \citet{Murphy2010} and \citet{Zhang2011} have proposed almost
the same variable selection methods based on the BIC criterion for the quadratic discriminant analysis. Let $\mathcal{S}=\{j_1,\cdots,j_m\}$ denote
a candidate model that contains the $X_{j_1}, \cdots, X_{j_m}$ as the relevant predictors, and $\mathcal{S}^c=\mathcal{S}_F\verb|\|S$, where $\mathcal{S}_F=\{1,2,\cdots,p\}$
is the set of all the candidate predictors. The BIC-type
criteria are based on the same assumptions:

(1) The reverent predictors $X_{\mathcal{S}}$ are the smallest set
of the candidate predictors which are sufficient for predicting the class label.
The assumption can be described by the following equality
\begin{equation}\label{eq:01}
P(G=k|X_{(\mathcal{S})},X_{(\mathcal{S}^c)})=P(G=k|X_{(\mathcal{S})}),
\end{equation}
where $X_{(\mathcal{S})}$ and $X_{(\mathcal{S}^c)}$
denote the subvector of the predictors corresponding
to the set $\mathcal{S}$ and $\mathcal{S}^c$.

It can be verified that (\ref{eq:01}) is equivalent to saying that
the irrelevant predictors are conditionally independent with $G$ given
the relevant predictors $X_{(\mathcal{S})}$, i.e.,
\begin{equation*}
P(X_{(\mathcal{S}^c)}|G,X_{(\mathcal{S})})=P(X_{(\mathcal{S}^c)}|X_{(\mathcal{S})}).
\end{equation*}

(2) All of the relevant and irrelevant predictors follow a jointly multivariate normal distribution given the class label. The conditional distribution of the relevant predictors given the class label is
\begin{equation*}
X_{(\mathcal{S})}|G=k\sim N\left(\mu_{k(\mathcal{S})},\Sigma_{k(\mathcal{S})}\right),
\end{equation*}
where $\mu_{k(\mathcal{S})}\in\mathbb{R}^{|\mathcal{S}|}$,
$\Sigma_{k(\mathcal{S})}\in\mathbb{R}^{(|\mathcal{S}|)\times(|\mathcal{S}|)}$
is a positive matrix, and $|\mathcal{S}|$ is the size of the set $\mathcal{S}$.
The conditional distribution of the irrelevant predictors
given the relevant predictors
is
\begin{equation*}
X_{(\mathcal{S}^c)}|X_{(\mathcal{S})}\sim N\left(\mu_{(\mathcal{S})}+B^T_{(\mathcal{S})}X_{\mathcal{S}},\Sigma_{\epsilon(\mathcal{S})}\right),
\end{equation*}
where $\mu_{(\mathcal{S})}\in\mathbb{R}^{p-|\mathcal{S}|}$,
$B_{(\mathcal{S})}\in\mathbb{R}^{(p-|\mathcal{S}|)\times|\mathcal{S}|}$,
and $\Sigma_{\epsilon(\mathcal{S})}\in\mathbb{R}^{(p-|\mathcal{S}|)\times(p-|\mathcal{S}|)}$
is a positive definite matrix.

Denote
\[\theta_{(\mathcal{S})}=\left\{\mu_{(\mathcal{S})},B_{(\mathcal{S})},\Sigma_{\epsilon(\mathcal{S})},
\mu_{k(\mathcal{S})},\Sigma_{k(\mathcal{S})},\pi_k,k=1,2\right\}.\]

Let$(x_i^T,g_i)^T\in\mathbb{R}^{p+1},i=1,2,\cdots,n$ denote the $n$
independent observations, where \\
$x_i=(x_{i1},x_{i2},\cdots,x_{ip})^T$
is the associated $p$-dimensional predictors collected from the $i$th subject.
If $\mathcal{S}$ is the smallest set of the relevant predictors,
the full likelihood function $\ell(\theta_\mathcal{S}|x_i,g_i,i=1,\cdots,n)$
can be written as
\begin{eqnarray*}
&&\ell(\theta_{(\mathcal{S})}|x_i,g_i,i=1,\cdots,n)\\
&=&\prod_{i=1}^nP(x_i,g_i|\theta_{(\mathcal{S})})\\
&=&\prod_{i=1}^nP(x_{i(\mathcal{S}^c)}|x_{(\mathcal{S})},\mu_{(\mathcal{S})},B_{(\mathcal{S})},\Sigma_{\epsilon(\mathcal{S})})
P(x_{i(\mathcal{S})}|g_i,\mu_{k(\mathcal{S})},\Sigma_{k(\mathcal{S})})P(g_i|\pi_k),
\end{eqnarray*}
where $X_{i(\mathcal{S})}$ and $X_{i(\mathcal{S}^c)}$
denote the subvector of the predictors collected from the $i$th subject corresponding
to the set $\mathcal{S}$ and $\mathcal{S}^c$.

Let $\hat\theta_{\mathcal{S}}$ be the maximum likelihood estimators.
The BIC proposed by \cite{Zhang2011} and
\cite{ Murphy2010} based on the full likelihood is defined as
\begin{equation*}
\text{BIC}(\mathcal{S})=-2\log\ell(\hat\theta_{(\mathcal{S})})+
\text{df}(\mathcal{S})\log n,
\end{equation*}
where $\text{df}(\mathcal{S})$ is the number of parameters needed
for the model with selected predictors $X_{(\mathcal{S})}$.

Even though the BIC criterion was proved to be consistent,
it is not applicable when the sample size $n_k$ is less than the dimension
$p$ of the predictors for any class $k=1,2$, and is clearly ill-posed if
$n_k<p$. When the sample size $n_k$ is less than the
dimension $p$, \citet{Fan2015} proposed
a two-step procedure for sparse QDA (IIS-SQDA).
For the two-class mixture Gaussian classification, the Bayes rule is an equivalent decision rule of the following form,
\begin{equation}\label{eq:02}
    Q(x) = \frac{1}{2}x^T(\Sigma_2^{-1}-\Sigma_1^{-1})x+x^T(\Sigma_1^{-1}\mu_1-\Sigma_2^{-1}\mu_2)+\zeta,
\end{equation}
where $\zeta$ is some constant depending only
on $\pi_k,\mu_k,\Sigma_k,k=1,2$. A new observation $x$
is predicted as the class $1$ if and only if $Q(x)>0$.
The first step in IIS-SQDA is to sparsify the support
$\Omega=\Sigma^{-1}_2-\Sigma^{-1}_1$ for interaction screening.
Two transformations based on the precision matrices
are used to find the interaction variables.
One regularization method was proposed for further selecting important interactions and main effects in the second step of IIS-SQDA.
Although IIS-SQDA is proved to enjoy the  sure screening property
in selecting the interactions and is close
to the the oracle classification in the performance of the misclassification, it
depends on the Gaussian mixture assumption and is not robust.

\section{Regularization methods and coordinate descent algorithm}
The approach to selecting variable proposed in this work is motivated by the
Bayes decision function (\ref{eq:02}) for the binary-class QDA.
Suppose we numerically code the class labels $g=1$
and $g=2$, respectively, as $y=1$ and $y=-1$.
We use the linear regression model where the predictors
are the main and interaction effects of the variables.
The coefficients of the linear regression model are estimated
by the following least squares
\begin{equation}\label{eq:03}
    (\hat\beta_0^{\text{ols}},\hat\beta_\odot^{\text{ols}},\hat\beta_\otimes^{\text{ols}})
    =\mathop{\arg\min}_{\beta_0, \beta_\odot, \beta_\otimes}\sum_{i=1}^n\left(
      y_i-\beta_0-x_i^T\beta_\odot-\tilde x_i^T\beta_\otimes
    \right)^2,
\end{equation}
where $\beta_0\in\mathbb{R}, \beta_\odot\in\mathbb{R}^p, \beta_\otimes\in\mathbb{R}^{\frac{p(p+1)}{2}}$,
and $\tilde x_{i}$ denotes the vector of all
interaction effects with the following form
\begin{equation*}
\tilde x_{i}
=(x^2_{i1},x_{i1}x_{i2},\cdots,x_{i1}x_{ip},x^2_{i2},x_{i2}x_{i3},\cdots,x^2_{ip})^T.
\end{equation*}
Denote $\beta=(\beta^T_0,\beta^T_\odot,\beta^T_\otimes)^T$.
For the linear regression model
in the variable selection problem, the classical regularized estimates of
the parameters $\beta$ are given by a penalized least squares
\begin{equation*}
    \hat\beta(\lambda)
    =\mathop{\arg\min}_{\beta_0, \beta_\odot, \beta_\otimes}\sum_{i=1}^n\left(
      y_i-\beta_0-x_i^T\beta_\odot-\tilde x_i^T\beta_\otimes
    \right)^2+P_{\lambda}(\beta),
\end{equation*}
where $\lambda$ is a tuning parameter(s) and controls the amount of regularization, and $P_{\lambda}(\beta)$ denotes a generic penalty function.

Some well-known regularization methods are lasso \citep{Tibshirani1996}, SCAD \citep{Fan2001}, elastic net \citep{Zou2005}, fused lasso \citep{Tibshirani2005}, grouped lasso \citep{Yuan2006}, adaptive lasso \citep{Zou2006}, MCP \citep{Zhang2010}, SICA \citep{Lv2009} and CAP \citep{Zhao2009}, among others.
To apply the group selection
for the overlapping patterns of the groups,
we propose to use the composite absolute penalty (CAP) \citep{Zhao2009},
which allows overlapping
patterns of the groups and different norms to be
combined in the penalty.
Therefore, we named our proposed resulting classifier the CAP-SQDA.
Let $\beta_\odot = (\beta_1,\beta_2,\cdots,\beta_p)^T$
and $\beta_\otimes = (\beta_{1,1},\beta_{1,2},\cdots,
\beta_{1,p},\beta_{2,2},\beta_{2,3},\cdots,\beta_{p,p})^T$.
For the linear regression model, the form of the CAP is
\begin{equation*}
P_\lambda(\beta)=\lambda\sum_{k=1}^{p}\sum_{l=k}^{p}\left[\alpha_{1,k,l}|\beta_{k,l}|+
                   \alpha_{2,k,l}\|v(\beta_k,\beta_l,\beta_{k,l})\|_{\gamma_{k,l}}\right],
\end{equation*}
where $\lambda$ is the tuning parameter, $\alpha_{1,k,l}>0$ and $\alpha_{2,k,l}>0$ are the weighted factors,
$\|\cdot\|_{\gamma}$ denotes the $L_\gamma$ norm,
and $v(\beta_k,\beta_l,\beta_{k,l})$ denotes the vector consisting
of $\beta_k,\beta_l,\beta_{k,l}$ with the following form:
\begin{equation*}
v(\beta_k,\beta_l,\beta_{k,l})=\left\{
\begin{aligned}
&(\beta_k,\beta_l,\beta_{k,l})^T,&k<l,\\
&(\beta_k,\beta_{k,k})^T,&k=l.
\end{aligned}
\right.
\end{equation*}

In our application, we keep the uniform
weights $\alpha_{1,k,l}=\alpha_1$,
$\alpha_{2,k,l} = \alpha_2$,
and uniform norms $\gamma_{k,l}=2$.
The penalty function in our proposed method is expressed
as follows
\begin{equation*}
P_{\lambda_1,\lambda_2}(\beta)=\sum_{k=1}^{p}\sum_{l=k}^{p}\left[\lambda_1|\beta_{k,l}|+
                   \lambda_{2}\|v(\beta_k,\beta_l,\beta_{k,l})\|_2\right],
\end{equation*}
where $\lambda_1$ and $\lambda_2$ are tuning parameters.

We suppose that an interaction effect should be added to
the regression model only after the corresponding main effects.
It means that the penalty for the interactions
should be larger than that for the main effects.
Thus we make the constraint ${\lambda_1}/{\lambda_2}>p$
for the two tuning parameters in the penalty function.
The constraint for the tuning parameters is also
identical to the IIS-SQDA \citep{Fan2015}.
Hence CAP-SQDA proposed in this work
is able to adaptively and automatically
choose between sparse QDA and sparse LDA
by using the penalized linear regression.

In optimizing the penalized linear regression, we always center
each predictor variable. When we center
all variables (including all predictors $x_i$ and $\tilde x_i$ and
all codes $y_i$ for
$i=1,2,\cdots,n$), the optimum value
of $\beta_0$ is $0$ for all values $\lambda_1$ and
$\lambda_2$. Then the optimization for CAP-SQDA can be
expressed in the more explicit form as
\begin{eqnarray}\label{eq:04}
    (\hat\beta_\odot,\hat\beta_\otimes)&=&\mathop{\arg\min}_{\beta_\odot, \beta_\otimes}\sum_{i=1}^n\left(
      y_i-x_i^T\beta_\odot-\tilde x_i^T\beta_\otimes
    \right)^2\\\nonumber
    &+&
    \sum_{k=1}^{p}\sum_{l=k}^{p}\left[\lambda_1|\beta_{k,l}|+
                   \lambda_{2}\|v(\beta_k,\beta_l,\beta_{k,l})\|_2\right].
\end{eqnarray}
\citet{Zhao2009} proposed
using the BLASSO algorithm to compute CAP estimates
in general. However, the BLASSO algorithm is tries to solve the whole solution path and so is only applicable for one
tuning parameter. There are two tuning parameters in our proposed optimization problem
(\ref{eq:04}). Therefore BLASSO is not appropriate for the optimization (\ref{eq:04}).
Cyclical coordinate descent methods are natural approaches for solving
convex problems $\ell_1$ and $\ell_2$ constraints. These
methods have been widely proposed for the lasso-type regularization problems,
including the classical sparse group lasso \citep{Friedman2008},
and the glmnet \citep{Friedman2010}.
We also use the coordinate descent method to solve the penalized
least squares problem (\ref{eq:04}).

The coordinate descent algorithm
for solving the optimization (\ref{eq:04})
can be converted to a general one-dimensional optimization
\begin{equation}\label{eq:05}
\mathop{\arg\min}_{\theta\in\mathcal{R}}a\theta^2+b\theta+c|\theta|+d\sum_{j=1}^s\sqrt{\theta^2+e_j},
\end{equation}
where $a\geq0,c\geq0,d\geq0,e_j>0$ and $s\in\{0,1,2,\cdots,p\}$.

If $|b|\leq c$, the minimizer is easily seen to be $\hat\theta=0$.
If $|b|>c$, the one-dimensional optimization (\ref{eq:05})
can be solved by Newton's type method or the {\bf optimize} function in the R
packages, which is a combination of golden section search and successive parabolic
interpolation. Specially, if $b>c$, the minimizer $\hat\theta$ lies in the interval
$\left(\frac{c-b}{2a},0\right)$; otherwise if $b<-c$,
the minimizer $\hat\theta\in\left(0,\frac{-c-b}{2a}\right)$.

To apply the one-dimensional optimization (\ref{eq:05})
in the CAP-SQDA, denote the following matrices
\begin{eqnarray*}
&&H=\left(\sum_{i=1}^nx^2_{i1},\sum_{i=1}^nx^2_{i2},\cdots,\sum_{i=1}^nx^2_{ip}\right)^T,\\
&&\widetilde H =
\left(\sum_{i=1}^n\tilde x^2_{i1},\sum_{i=1}^n\tilde x^2_{i2},\cdots,\sum_{i=1}^n\tilde x^2_{i\tilde p}\right)^T,\\
&&\mathbb{X} = (x_1,x_2,\cdots,x_n)^T\in\mathbb{R}^{n\times p},\\
&& \widetilde{\mathbb{X}} = (\tilde x_1,\tilde x_2,\cdots,\tilde x_n)^T\in\mathbb{R}^{n\times \tilde p},\\
&&\mathbb{Y}=(y_1,y_2,\cdots,y_n)^T.
\end{eqnarray*}
where $\tilde p = p(p+1)/2$.
We also need to compute
the following products in the coordinate descent algorithm
\begin{eqnarray*}
&&C = \mathbb{X}^T\mathbb{Y}, \quad\tilde C = \widetilde {\mathbb{X}}^T \mathbb{Y},\\
&&G = \mathbb{X}^T\mathbb{X}, \quad\tilde G = \widetilde {\mathbb{X}}^T\widetilde {\mathbb{X}}, \quad B = \mathbb{X}^T\widetilde{\mathbb{X}}.
\end{eqnarray*}

For the update of the main effect parameters $\beta_k$, the parameters in Equation \eqref{eq:04}
$a,b,c,d,e_j$ can be calculated as
\begin{eqnarray*}
&&a = H, \quad L = \{l\neq k:I(\beta^2_l+\beta^2_{k,l}>0)\}\cup
\{k:\beta^2_{k,k}>0\},\\
&&s = |L|, \quad c = (p-s)\lambda_2,\quad d = \lambda_2,\quad e_j=I(L(j)\neq k)\beta^2_{L(j)}+\beta^2_{k,L(j)},\\
&&b = 2 \left(\beta^T_{\odot\backslash k}G_{k, \backslash k}+\beta^T_\otimes B_k-C_k
\right),
\end{eqnarray*}
where $I(\cdot)$ denotes the indictor function,
$\beta_{\odot\backslash k}$ the subvector of $\beta_\odot$ removing the $k$th element,
$G_{k,\backslash k}$ the $k$th row of $G$ with the $k$th element removed,
$B_k$ the $k$th row of $B$, and
$C_k$ the $k$th element of $C$.

For the update of the interaction effect parameter $\beta_{k,l}$, the parameters in Equation \eqref{eq:05}
$a,b,c,d,e_j$ can be calculated as
\begin{eqnarray*}
&&m = (k-1)p-(k-1)(k-2)/2+(l-k+1),a = \widetilde H, \\
&&s = I(k\neq l)I(\beta_k^2+\beta_l^2>0)+I(k=l)I(\beta_k^2>0),\\
&&e_1=I(k\neq l)(\beta_k^2+\beta_l^2>0)+I(k=l)(\beta_k^2>0),\\
&&c = \lambda_1+\lambda_2I(s=0),d=\lambda_2I(s=1),\\
&&d = 2(\beta_\odot^T B^m+\beta^T_{\otimes\backslash m}\tilde G_{m,\backslash m}
      -\tilde C_m),
\end{eqnarray*}
where $B^m$ denotes the $m$th column of $B$.
This leads to the following algorithm:

{\bf Step 1}: Start with $(\hat\beta_\odot,\hat\beta_\otimes)^T=(\beta_\odot^{(0)},\beta_\otimes^{(0)})$.

{\bf Step 2}: For the updated estimate of $\beta_k$ in the $t$th loop, fix $\beta_l,l\neq k$ and $\beta_{k,l}$, and calculate $a,b,c,d,e_j$; if $|b|<c$, set $\beta_k^{(t+1)}=0$; otherwise minimize the optimization (\ref{eq:05}) and obtain the update $\beta_k^{(t+1)}$ of $\beta_k$. Update all the parameters $\beta_k,1\leq k\leq p$
and $\beta_{k,l},1\leq k\leq l\leq p$ in order.

{\bf Step 3}: Iterate the entire step (2) over
$t=1,2,\cdots$ until convergence.

Let $\lambda = (\lambda_1,\lambda_2)$, and denote $\hat\beta_0,\hat\beta_\odot(\lambda),\hat\beta_\otimes(\lambda)$ as the estimates
of the parameters $\beta_0,\beta_\odot,\beta_\otimes$ respectively
by solving the penalized linear regression problem. Then
the classification rule is to assign a new observation $z\in\mathcal{R}^p$ to class $1$
if and only if $\hat\beta_0+ z^T\hat\beta_\odot(\lambda)+ \tilde z^T\beta_\otimes(\lambda)>0$
where $\tilde z$ is the corresponding interaction effects to $z$.
In practice, we need to select a good tuning parameter such that the
misclassification error is as small as possible.
Five cross-validation (CV) is a popular method
for tuning, and hence we use it here. Note that there are two tuning parameters in the CAP-SQDA, so the detail of CV can be referred to as the elastic net \citep{Zou2005}.
The Lasso-type estimates are generally biased.
Therefore, we suggest that OLS is used in the CAP-QDA
if the dimension of the active main and interaction
effects in the penalized linear regression  is smaller than the sample size.

The active sets for the main and interaction effects
are
\begin{eqnarray*}
&&S_1=\{k:\hat\beta_k(\lambda)\neq 0\},\\
&&S_2=\{(k-1)p-(k-1)(k-2)/2+(l-k+1):\hat\beta_{k,l}(\lambda)\neq 0\}.
\end{eqnarray*}
If $|S_1|+|S_2|<n$, the OLS estimates of $\beta_{\odot S_1}$
and $\beta_{\otimes S_2}$ have the following form
\begin{equation*}
(\hat\beta_{\odot S_1}^{\text{ols}},\hat\beta_{\otimes S_2}^{\text{ols}})=
\left[(x_{S_1},\tilde x_{S_2})^T(x_{S_1},\tilde x_{S_2})\right]^{-1}(x_{S_1},\tilde x_{S_2})^Ty.
\end{equation*}
Then if the dimension of the selected effects is smaller than
the sample size, the sparse QDA classifier is defined as follows:
assigning the new observation $z$ and the corresponding
interactions $\tilde z$ to class $1$ if
\begin{equation*}
\hat\beta_0+z^T_{S_1}\hat\beta_{\odot S_1}^{\text{ols}}+
\tilde z_{S_2}^T\hat\beta_{\otimes S_2}^{\text{ols}}>0.
\end{equation*}

\section{Simulation studies}
In this section, a number of simulations are conducted to compare the
performance of the proposed methods with
the two methods based on the full likelihood
BIC: the backward procedure presented in
\citet{Zhang2011}, denoted by  $\text{BIC}_{b}$ method,
and the forward-backward procedure proposed in
\citet{Murphy2010}, denoted by $\text{BIC}_{fb}$  method.
Five simulation experiments are considered.
In the first three experiments named Model $1$ - Model $3$, the predictors
are generated from the multivariate normal
distributions, and the predictors in the other two experiments
named Model $4$ and Model $5$ are not multivariate normally distributed  random variables.
In each simulation experiment, we consider low-dimensional settings with $p=20$
and high-dimensional settings with $p=100,200$.

For each simulation experiment setting,
$50$ observations for each class generated from the true model are served as the training data
while $5000$ extra independent observations for each class
are served as the testing data.
For comparison, we consider five performance measures
including misclassification rate (MR), the numbers
of irrelevant main effects (FP.main) and irrelevant interaction
effects (FP.inter) falsely included in the classification rule,
and the numbers of relevant main effects (FN.main) and interaction effects
(FN.inter) falsely excluded in the classification rule.
The five performance measures are the same as the classification and variable selection performances employed in \citet{Fan2015}.

(1) Model $1$: The relevant predictors are ${\bf X}_\mathcal{S}=\left\{X_1,X_2\right\}$.
For the first class,
the mean vector and covariance matrix
are ${\bf\mu}_{1,\mathcal{S}}=(2.5,-1)^T$
and ${\bf\Sigma}_{1,\mathcal{S}}=[1,0;0,1]\in\mathbb{R}^2$, respectively;
for the second class, they are
${\bf\mu}_{2,\mathcal{S}}=(-0.5,0)^T$
and ${\bf\Sigma}_{2,\mathcal{S}}=[3,1;1,3]\in\mathbb{R}^2$, respectively.
The
remaining $p-2$ variables are independently and independently generated as $N(u,1)$ , where
$u$ is generated from $U[0,1]$.

This model is borrowed from \cite{Zhang2011}.
There are two main effects and three interaction terms
in the Bayes rules for Model $1$.
There is small difference between the two
covariance matrices of the predictors for two classes. It means that
the interaction effects are weak in Model $1$.

\begin{table}[!htp]
\caption{Performance measures of
different classification methods for Model $1$ \label{Tab:01}}
\centering
\begin{tabular}{llllllllllllllllllllllllllll}
\toprule
$p$&  Method&           MR(\%)&  FP.main& FP.inter  &FN.main &FN.inter&\\
\midrule
20&
$\text{BIC}_{b}$&   6.68(0.57)&0.06(0.23)&0.14(0.56)&0.77(0.42)&1.54(0.84)&\\
&$\text{BIC}_{fb}$& 6.67(0.58)&0.06(0.28)& 0.14(0.56)&0.75(0.43)&1.50(0.87)&\\
&   IIS-SQDA&       6.92(1.69)&0.95(2.55)&0.84(0.63)&0.11(0.39)&2.98(0.14)&\\
&   CAP-SQDA&       6.54(0.48)&1.04(2.73)&0.48(1.46)&0.52(0.50)&2.65(0.55)\\
&   OLS-SQDA&       6.91(1.05)&1.04(2.73)&0.48(1.46)&0.52(0.50)&2.65(0.55)\\
&   ORACLE&         6.44(0.25)&    0(0)&   0(0)&    0(0)&  0(0)&\\[4pt]
100&
$\text{BIC}_{fb}$&  7.05(0.99)&0.36(0.52)& 0.81(1.22)&0.68(0.46)&  1.36(0.93)&\\
&   IIS-SQDA&       7.00(2.21)&1.31(4.94)&0.95(0.88)&0.06(0.23)&   2.98(0.14)&\\
&   CAP-SQDA&       6.54(0.49)&2.06(3.82)&0.50(1.23)&0.46(0.50)&2.93(0.26)\\
&   OLS-SQDA&       6.76(0.74)&2.06(3.82)&0.50(1.23)&0.46(0.50)&2.93(0.26)\\
&   ORACLE&         6.40(0.24)&   0(0)&   0(0)&    0(0)&  0(0)&\\[4pt]
200&
$\text{BIC}_{fb}$&  7.13(0.97)&0.49(0.65)& 1.11(1.57)&0.74(0.44)&  1.48(0.88)&\\
&   IIS-SQDA&       7.07(2.48)&1.05(3.15)&0.87(0.33)&0.18(0.38)&   2.99(0.10)&\\
&   CAP-SQDA&       6.55(0.47)&0.65(1.44)&0.51(1.28)&0.60(0.49)&2.67(0.47)\\
&   OLS-SQDA&       7.13(1.15)&0.65(1.44)&0.51(1.28)&0.60(0.49)&2.67(0.47)\\
&   ORACLE&         6.38(0.23)&   0(0)&   0(0)&    0(0)&  0(0)&\\
\bottomrule
\end{tabular}
\end{table}

Table \ref{Tab:01} presents the variable selection and classification results
for Model $1$. It can be seen from the table
that different methods exhibit similar
performance in MR. CAP-SQDA has the best classification performance
in all settings $p=20,100$ and $200$ although
the methods BIC and IIS-SQDA are consistent with
the assumptions of Model $1$.
BIC has the smallest values of
FP.main whereas IIS-SQDA has the
smallest values of FN.main in the main effect
selection under all settings.
CAP-SQDA gives smaller FN.mains than BIC.
CAP-SQDA has the smallest values of
FP.inter in the situations $p=100,200$
whereas BIC has the smallest values of
FN.inter across all settings
in terms of interaction selection.
Both of IIS-SQDA and CAP-SQDA
have poor interaction selection
performances which can be shown
that the values of FP.inter are
approximately equal to $3$.
The reason is that model $1$ is similar to
a LDA model. BIC criterion is
a method for variable selection rather
than effect selection essentially, therefore,
presents the best performance in
terms of FN.inter.
The results in Table \ref{Tab:01}
demonstrate that our proposed method
CAP-SQDA can effectively select important
effects and conduct classification simultaneously
for Model $1$ under all dimensional settings.

(2) Model $2$: The mean vectors of all variables are  $\mu_1=(0,0,\cdots)^T$ and
$\mu_2=(0.6,0.8,0,\cdots)^T$ for the two classes. The precision matrix for the first class
is $\Omega_1=I_{p\times p}$, and $\Omega_2=\Omega_1+\Omega$ for the other class,
where $\Omega$ is a sparse matrix with
$\Omega_{3,3}=\Omega_{4,4}=\Omega_{5,5}=-0.6$ and
$\Omega_{3,4}=\Omega_{3,5}=\Omega_{4,5}=-0.15$.
The other three nonzero entries in the lower triangle of $\Omega$
are determined by symmetry.
This model is borrowed from \cite{Fan2015}.
There are two main effects and six interaction terms in the Bayes rules
for Model $2$. The relevant
predictors are ${\bf X}_\mathcal{S}=\left\{X_1,X_2,\cdots,X_5\right\}$.

\begin{table}[!htp]
\caption{Performance measures of
different classification methods for Model $2$ \label{Tab:02}}
\centering
\begin{tabular}{lllllllllllllllllllllllllllllll}
\toprule
$p$&  Method&           MR(\%)&  FP.main& FP.inter  &FN.main &FN.inter&\\
\midrule
20&
$\text{BIC}_b$&     26.73(2.38)&2.06(0.62)&1.49(1.74)&1.53(0.52)&2.78(1.49)&\\
&$\text{BIC}_{fb}$& 26.45(2.27)&2.10(0.57)&1.48(1.69)&1.52(0.52)& 2.66(1.49)&\\
&   IIS-SQDA&       24.30(3.08)&3.39(4.70)&0.88(1.83)& 0.18(0.43)&1.33(1.55)&\\
&   CAP-SQDA&       24.91(2.44)&2.34(4.11)&0.60(1.04)&0.67(0.80)&2.93(0.57)\\
&   OLS-SQDA&       26.39(2.64)&5.78(7.36)&2.11(5.65)&0.54(0.75)&2.91(0.77)\\
&   ORACLE&         18.63(0.38)&  0(0)&   0(0)&    0(0)&  0(0)&\\[4pt]
100&
$\text{BIC}_{fb}$&  26.41(2.45)&2.11(0.60)&2.12(1.79)&1.37(0.56)&2.83(1.33)&\\
&   IIS-SQDA&       27.21(4.65)&5.54(9.10)&0.26(0.62)& 0.34(0.55)&2.48(1.52)&\\
&   CAP-SQDA&       26.23(2.89)&2.59(3.19)&0.58(1.62)&0.54(0.67)&3.37(0.79)\\
&   OLS-SQDA&       28.30(3.34)&1.93(1.77)&0.16(0.56)&0.62(0.74)&3.23(0.77)\\
&   ORACLE&         18.62(0.42)&  0(0)&   0(0)&    0(0)&  0(0)&\\[4pt]
200&
$\text{BIC}_{fb}$&  27.04(3.05)&2.27(0.71)&2.34(1.86)&1.45(0.65)&2.75(0.72)&\\
&   IIS-SQDA&       28.31(4.69)&6.05(10.80)&0.37(0.84)& 0.45(0.55)&2.96(1.37)&\\
&   CAP-SQDA&       25.87(2.84)&1.92(2.76)&0.53(1.23)&0.86(0.63)&2.01(0.69)\\
&   OLS-SQDA&       28.27(3.70)&1.93(1.77)&0.16(0.56)&0.62(0.74)&3.23(0.77)\\
&   ORACLE&         18.60(0.37)&  0(0)&   0(0)&    0(0)&  0(0)&\\
\bottomrule
\end{tabular}
\end{table}

(3) Model $3$: The mean vectors are $\mu_1=(0,0,\cdots)^T$ and
$\mu_2=(0.6,0.8,0.6,0.8,0,\cdots)^T$ for the two classes. The precision matrix is $\Omega_1=I_{p\times p}$ for the first class, and $\Omega_2=\Omega_1+\Omega$ for
the second class, where $\Omega_{1,1}=\Omega_{2,2}=-0.6$ and
$\Omega_{1,2}=\Omega_{2,1}=-0.15$.
This model is consistent with the effects assumption in our proposal.
There are four main effects and two interaction terms in the Bayes rules.
The relevant predictors are ${\bf X}_\mathcal{S}=\left\{X_1,X_2,X_3,X_4\right\}$.
\begin{table}[!htp]
\vskip2mm
\caption{Performance measures of
different classification methods for Model 3 \label{Tab:03}}
\centering
\begin{tabular}{llllllllllllllllllllllll}
\toprule
$p$&  Method&           MR(\%)&  FP.main& FP.inter  &FN.main &FN.inter&\\
\midrule
20&
$\text{BIC}_b$&     30.67(3.06)& 0.03(0.17)&1.40(1.44)&2.05(0.62)&1.23(1.08)&\\
&$\text{BIC}_{fb}$& 30.26(2.85)&0.02(0.14)& 1.34(1.28)&2.00(0.56)&1.12(1.05)&\\
&   IIS-SQDA&       27.74(3.24)&2.50(3.58)& 0.51(1.25)&1.10(0.84)&0.73(0.87)&\\
&   CAP-SQDA&       28.47(3.20)&2.43(4.58)&0.10(0.33)&1.54(1.26)&1.33(0.65)\\
&   OLS-SQDA&       29.03(3.37)&2.43(4.58)&0.10(0.33)&1.54(1.26)&1.33(0.65)\\
&   ORACLE&         22.68(0.35)&0(0)&   0(0)&    0(0)&  0(0)&\\[4pt]
100&
$\text{BIC}_{fb}$&  30.35(3.58)&0.14(0.37)&2.07(1.62)&1.92(0.59)& 1.32(1.01)&\\
&   IIS-SQDA&       30.73(4.48)&6.10(8.74)&0.50(0.91)& 1.14(0.89)&1.40(0.95)&\\
&   CAP-SQDA&       29.18(2.87)&3.77(6.95)&0.03(0.17)&1.62(0.97)&1.62(0.78)\\
&   OLS-SQDA&       30.78(3.87)&3.77(6.95)&0.03(0.17)&1.62(0.97)&1.62(0.78)\\
&   ORACLE&         22.67(0.45)&0(0)&   0(0)&    0(0)&  0(0)&\\[4pt]
200&
$\text{BIC}_{fb}$&  32.01(4.00)&0.36(0.54)&2.21(1.72)&2.11(0.52)& 1.40(0.98)&\\
&   IIS-SQDA&       33.82(5.48)&10.06(16.02)&0.42(0.72)& 1.43(0.90)&1.55(0.96)&\\
&   CAP-SQDA&       29.90(2.95)&2.69(3.81)&0.06(0.31)&1.89(0.97)&1.28(0.53)\\
&   OLS-SQDA&       31.02(4.22)&2.69(3.81)&0.06(0.31)&1.89(0.97)&1.28(0.53)\\
&   ORACLE&         22.70(0.41)&0(0)&   0(0)&    0(0)&  0(0)&\\
\bottomrule
\end{tabular}
\end{table}

(4) Model $4$: Assume that part of the variables is not
Gaussianly distributed for each class. Suppose also that the first two variables are relevant.
For the first class, $X_1$ and $X_2$ are independently generated
from $1 -\chi^2(1)$, where $\chi^2(1)$ is the chi-squared distribution with
one degree of freedom. For the second class, $X_1$ and $X_2$
are independently generated from,
$1.2-\sqrt{3}*\chi^2(1)$ and $1.6-\sqrt{3}*\chi^2(1)$, respectively.
The next three variables are generated from the following rules
 \begin{eqnarray*}
 &X_3=b_{11}+b_{12}X_1+\chi^2(1),\\
 &X_4=b_{21}+b_{22}X_2+\chi^2(1),\\
 &X_5=b_{31}+b_{32}X_1+b_{33}X_2+\chi^2(1),
 \end{eqnarray*}
 where $b_{\cdot\cdot}$ is $U[-1,1]$.
 The next $p/2-5$ variables are independently generated
 from the standard normal distribution.
 The remaining $p/2$ variables are independently generated
 from $\chi^2(1)$.

(5) Model $5$: The first five variables are generated as in Model $4$.
The remaining $p-5$ are irrelevant.
The next $p/2-5$ variables are independently generated from
$N(\tilde\mu,1)$, and the remaining $p/2$ variables are independently generated from
Beta distribution $B(\nu,0.5)$, where $\tilde\mu$ and $\nu$ are $U(0,1)$
and $U(1,5)$ random variables, respectively.

Table \ref{Tab:04} summarizes the classification results for Model $4$ and Model $5$.
We not only give the misclassification rates, but also
report the values of MR of CAP-SQDA minus the ones of BIC and IIS-SQDA(MRM).
According to Table \ref{Tab:04}, although the misclassification rates of all methods in models $4-5$ are obviously larger than those of the ORACLE classifier
when the assumption of mixed gaussian distribution
for all variables does not hold,
CAP-SQDA exhibits the best
performance in terms of MR across all settings.
In some settings such as $p=100,200$ for Model $5$, MRs of
CAP-SQDA are significantly smaller than the ones of BIC and IIS-SQDA.
The results demonstrate that our proposed method is robust for QDA in classification.

\begin{table}[!htp]
\vskip2mm
\caption{Performance measures of different classification methods for Model $4$ and Model $5$
 \label{Tab:04}}
\centering
\begin{tabular}{lllllllllllllllllll}
\toprule
&       &\multicolumn{2}{c}{model 4}& &\multicolumn{2}{c}{model 5}\\
         \cline{3-4}\cline{6-7}
 $ p $  &Method                   &MR(\%)       &MRM(\%) &&MR(\%)       &MRM(\%)\\
\midrule
20    & $ \text{BIC}_b $    &24.96(10.26)   &7.76(10.03)    &&19.44(8.62)  &4.86(9.01)\\
       & $ \text{BIC}_{fb} $& 24.53(10.35)   &7.33(9.89)    &&19.31(8.53)  &4.73(8.82)\\
       &IIS-SQDA            & 23.53(7.30)    &6.32(6.95)    &&22.01(7.66)  &7.42(7.76)\\
      &CAP-SQDA             &17.20(3.32)     &--            &&14.58(2.75)  &-- \\
     &OLS-SQDA              &18.78(4.17)     &--            &&14.96(3.16)  &-- \\
    & ORACLE                 & 4.63(0.18)    &--            &&4.62(0.18)   &-- \\[4pt]
100   & $ \text{BIC}_{b} $  & 33.30(9.43)   &9.33(10.71)   &&23.52(8.64)  &7.82(8.22)\\
       &IIS-SQDA            &29.79(8.02)    & 5.82(8.89)   &&27.91(8.54)  &12.21(8.35) \\
      &CAP-SQDA             &23.96(6.40)    &--            &&15.70(2.66)  &-- \\
     &OLS-SQDA              &24.56(6.73)    &--            &&16.72(3.24)  &--\\
    & ORACLE                & 4.63(0.18)    &--            &&4.63(0.18)   &--\\[4pt]
    200   & $ \text{BIC}_{b}$&39.10(8.01)   &12.95(9.06)   &&27.31(7.54)  &10.73(7.87)\\
       &IIS-SQDA            &32.33(9.19)    &6.18(9.36)   &&31.22(8.85)  &14.65(8.64) \\
      &CAP-SQDA             &26.15(6.42)    &--           &&16.57(2.86)  &--  \\
     &OLS-SQDA              &27.01(7.26)    &--            &&17.36(3.83)  &-- \\
    & ORACLE                &4.63(0.18)     &--           &&4.63(0.18)   &-- \\
    \bottomrule
\end{tabular}
\end{table}

\section{Application}
\subsection{parkinson dataset}
We apply the classification methods to the parkinson dataset shared in
UCI in $2008$ \citep{Little2007}. This dataset is composed of a range of biomedical voice measurements from normal people and Parkinson's disease (PD) patients. There are $p=22$ predictors in this dataset. The main aim is to discriminate healthy people from those with PD.
There are $n_1=147$ for the PD and $n_2=48$.
We randomly split $195$ samples into a training set consisting of $73$
samples from PD and $24$ samples from the healthy.
For each split, we applied five different methods to the training data and then calculated the classification error using the test data.
The tuning parameters are selected
via the five-fold cross validation. We repeated the random splitting for $100$ times. The means and standard errors of classification errors and model sizes for different classification methods are summarized in Table \ref{Tab:05}.

FULL method has the worst performance in the classification.
It means that the variable selection is necessary
for the analysis of the parkinson dataset.
$\text{BIC}_{b}$ method and $\text{BIC}_{fb}$ method select on average
$10.63$ and $7.88$ variables, respectively.
IIS-SQDA method has the smallest MR, but select
$11.20$ variables and $20.94$ effects.
Our proposed method selects the smallest numbers
of variables and effects and and achieves very close
classification accuracy compared with the IIS-SQDA method.
\begin{table}[!htp]
\caption{Analysis of the parkinson data over $100$ random splits. \label{Tab:05}}
\centering
\begin{tabular}{llllllllllllllllllllllllllll}
\toprule
Method&           MR(\%)&  Variable& Main&  Interaction&   All&\\
\midrule
FULL&               22.51(1.92)&-- &--      &--         &--         &\\
$\text{BIC}_{b}$&   18.01(3.99)& 10.63(1.81)&--         &--         &--      &\\
$\text{BIC}_{fb}$&  21.36(4.33)& 7.88(2.00)&--          &--         &--      &\\
IIS-SQDA&           15.78(3.91)& 11.20(4.19)&9.92(4.33)&11.02(8.60)&20.94(10.59)&\\
   CAP-SQDA&        15.94(2.72)& 4.44(1.70)& 2.86(1.93)&2.12(0.79) &4.98(2.14)&\\
   OLS-SQDA&        16.11(3.04)& 4.44(1.70)& 2.86(1.93)&2.12(0.79)&4.98(2.14)&\\
\bottomrule
\end{tabular}
\begin{tablenotes}
\footnotesize
\item \hskip 1em{\bf NOTE:} FULL represents the QDA without variable selection.
\end{tablenotes}
\end{table}

\subsection{Breast cancer dataset}
The breast cancer dataset consists of the gene expressions from
$77$ patients, originally studied in \citet{Van2002}. The goal is to
predict whether a female breast cancer patient relapses
from gene expression data. The dataset contains a
total of $78$ samples, with $44$ of them in the good prognosis group and
$34$ of them in the poor prognosis group.
Since there are some missing values with
one patient in the poor prognosis group,
it was removed in the study \cite{Fan2015}.
Same as \cite{Fan2015}, we use the $p=231$ genes in \citet{Van2002}
and randomly split the $77$ samples into a training set and a test set.
$26$ samples from the good prognosis group
and $19$ samples from the poor prognosis group are randomly selected
in the  training set randomly. We apply the classification methods and
the results are summarized in Table \ref{Tab:06}.

For the analysis of the breast cancer dataset, $\text{BIC}_{b}$ method and $\text{BIC}_{fb}$ method select on average
$10.63$ and $7.88$ variables, respectively.
IIS-SQDA method has the lower MR, but select
$11.20$ variables and $20.94$ effects.
Our proposed method selects the smallest number
of variables and effects and achieves high
classification accuracy.
$BIC_{fb}$ selects the smallest variables, but has the largest
MR. IIS-SQDA method achieves the lowest MN, but selects
the largest number of variables and effects.
where both of the numbers of the Main effects and All
effects are larger than the sample size of the training set.
Our proposal method misclassifies $2.46$ clinical outcomes more than
IIS-method, whereas it selects nearly half the number of effects fewer than the IIS-method. Both our proposal method and the IIS-method select very fewer interaction effects.
It means that a sparse LDA is suitable for the breast cancer dataset.
In the study of \citet{Fan2015}, the penalized logistic regression analysis
with the main effects only also has high classification accuracy
with $\text{MN}=6.95$.
It demonstrate that our proposal method can adaptively and automatically
choose between the sparse LDA and the sparse QDA.

\begin{table}[!htp]
\caption{Analysis of the breast cancer data over $100$ random splits \label{Tab:06}}
\centering
\begin{tabular}{llllllllllllllllllllllllllll}
\toprule
Method&           MN&   Main&  Interaction&   All&\\
\midrule
$\text{BIC}_{fb}$&   11.57(2.30)&7.72(3.26)&--  &--   &           &\\
   IIS-SQDA&         6.39(2.46) &47.77(11.60)&3.03(3.20) &50.80(13.10)&\\
   CAP-SQDA&         8.85(2.24)&21.42(9.07)&1.06(1.48)  &22.48(10.06)&\\
   OLS-SQDA&         9.25(2.34)&21.42(9.07)&1.06(1.48)  &22.48(10.06)&\\
\bottomrule
\end{tabular}
\begin{tablenotes}
\footnotesize
\item \hskip 3.5em {\bf NOTE:} MN represents the misclassification number.
\end{tablenotes}
\end{table}

\section{Conclusion}
In this paper we propose a penalized linear regression,named CAP-SQDA,
for quadratic discriminant analysis with two classes, and develop
a coordinate descent algorithm to solve the penalized least-squares problem.
The proposed procedure first transform the sparse QDA problem to a penalized sparse ordinary least squares optimization by using composite absolute penalty, and apply main effect and interaction selection through regularization.
The efficiency and robustness of CAP-SQDA have been demonstrated through simulation
studies and real data analysis through comparison it with other methods.
For real datasets CAP-SQDA usually selects much few variables while achieves high classification accuracy.

In the future study, it would be interesting to generalize the proposed method to problems for quadratic discriminant analysis of multi-class classification. The key
of the variable selection for the multi-class quadratic discriminant analysis
is to propose a new composite penalty. In addition, developing an efficient
computing method is the need of CAP-SQDA for ultrahigh-dimensional data analysis.

\section*{Acknowledgements}

 We would like to thank Dr. Yinfei Kong for providing the code of IIS-SQDA and the breast cancer data.

\label{lastpage}

\end{document}